\documentclass[journal]{IEEEtran}
\usepackage{algorithm}
\usepackage{algorithmic}
\usepackage{amsfonts,amssymb}
\usepackage{graphicx}
\usepackage{cite}
\usepackage{booktabs}
\usepackage{subfigure}
\usepackage{amsfonts,amssymb}
\usepackage{textcomp}
\usepackage{amsmath,amssymb,amsfonts}
\usepackage[justification=centering]{caption}
\usepackage{hyperref}
\usepackage{soul}
\soulregister\cite7 
\soulregister\citep7 
\soulregister\citet7 
\soulregister\ref7 
\soulregister\pageref7 
\usepackage[T1]{fontenc}
\usepackage[utf8]{inputenc}
\usepackage[margin=0.5in, headsep=10pt]{geometry}
\usepackage{fancyhdr}
\title{Smart Train Operation Algorithms based on Expert Knowledge and Reinforcement Learning}

\usepackage{lipsum}

\begin{document}

\author{Kaichen~Zhou,~\IEEEmembership{Student Member,~IEEE,}
        Shiji~Song,~\IEEEmembership{Member,~IEEE,}
        Anke~Xue,~\IEEEmembership{Member,~IEEE}
        Keyou~You,~\IEEEmembership{Member,~IEEE}
        and~Hui~Wu,~\IEEEmembership{Student Member,~IEEE}
    
\thanks{This work is funded by the key project of the National Natural Science Foundation of China under Grant 61936009, and is funded by the major research plan of the National Major Research Program under Grant 2018AAA0101604. This work was done when the first author was with Department of Automation, Tsinghua University, under the supervision of Prof.Shiji Song and Dr.Keyou You.}

\thanks{K. Zhou is with Department of Computer Science, University of Oxford, Oxford OX1 2JD, UK (e-mail: ruizhou2019@hotmail.com).}
\thanks{S. Song, K. You and H. Wu are with Department of Automation and BNRist, Tsinghua University, Beijing 100084, China (e-mail: shijis@mail.tsinghua.edu.cn; youky@tsinghua.edu.cn; wuhui115199@163.com).}
\thanks{A. Xue is with the Key Laboratory for IOT and Information Fusion Technology of Zhejiang, Institute of Information and Control, Hangzhou Dianzi University, Hangzhou 310018, China (e-mail: akxue@hdu.edu.cn).}}
\maketitle
\thispagestyle{fancy}

\begin{abstract}
During recent decades, the automatic train operation (ATO) system has been gradually adopted in many subway systems for its low-cost and intelligence.  This paper proposes two smart train operation algorithms by integrating the expert knowledge with reinforcement learning algorithms. Compared with previous works, the proposed algorithms can realize the control of continuous action for the subway system and optimize multiple critical objectives without using an offline speed profile. Firstly, through learning historical data of experienced subway drivers, we extract the expert knowledge rules and build inference methods to guarantee the riding comfort, the punctuality, and the safety of the subway system. Then we develop two algorithms for optimizing the energy efficiency of train operation. One is the smart train operation (STO) algorithm based on deep deterministic policy gradient named (STOD) and the other is the smart train operation algorithm based on normalized advantage function (STON). Finally, we verify the performance of proposed algorithms via some numerical simulations with the real field data from the Yizhuang Line of the Beijing Subway and illustrate that the developed smart train operation algorithm are better than expert manual driving and existing ATO algorithms in terms of energy efficiency. Moreover, STOD and STON can adapt to different trip times and different resistance conditions.
\end{abstract}


\IEEEpeerreviewmaketitle
\section{Introduction}
With the deterioration of modern urban traffic problems and energy problems, the urban subway is getting more and more attention due to its advantages on safety, punctuality, and energy efficiency \cite{yin2016smart}. Since January 9th, 1863, the first subway has started operation from Paddington to Farringdon, and until 2015, more than 150 cities were hosting approximately 160 subway systems around the world. 

\par Meanwhile, with the acceleration of the modernization process of the subway, the automatic train operation system has been used in many places to replace manual driving for its low-cost and intelligence. The ATO system first generates the target speed curve based on various requirements under both the train condition and the railway condition, and then sends a speed control command to control the train to track the generated target speed curve \cite{corman2015review}. Thus, the ATO system has a direct influence on the train's trajectory. And improving the performance of the ATO system has become a focus in the field of the transportation system. In most cases, research on the urban subway system can be divided into two parts: the energy-efficient train operation committed to designing an off-line optimized train trajectory, and the automatic train tracking method to track the real-time train speed-distance profile.

\par During recent years, lot of studies are devoted to designing an off-line optimized train trajectory to improve the energy efficiency. For example, Albrecht \emph{et al.} used a comprehensive perturbation analysis to show that a key local energy function is strictly convex with a unique minimum, and thus proved that the optimal switching points are uniquely defined for each steep section \cite{albrecht2013energy}. 
Besides, the train operation problem also involves many other aspects, such as riding comfort and punctuality. In order to minimize the total travel time of passengers and the energy consumption of the train operation, Wang \emph{et al.} proposed a new iterative convex programming (ICP) approach to obtain the optimal departure times, running times and dwell times, to solve the train scheduling problem \cite{wang2015efficient}. Focusing on the energy-saving and the service quality, Yang \emph{et al.} formulated a two-objective integer programming model with headway time and dwell time control to find the optimal solution by designing a genetic algorithm with binary encoding \cite{yang2014two}. Considering the energy consumption and trip time as the main objectives of optimization, Wei \emph{et al.} developed a multi-objective optimization model for the speed trajectory by using optimal speed trajectory searching strategies under different track characteristics \cite{shangguan2015multiobjective}. Moreover, the parameters of railways are not always constant. With the consideration of variable gradients and arbitrary speed limits, Khmelnitsky \emph{et al.} constructed a numerical algorithm to obtain the optimal velocity profile \cite{khmelnitsky2000optimal}. With the development of artificial intelligence, many intelligent algorithms have recently been used in the train operation. Açıkbaş \emph{et al.} implemented artificial neural networks with a genetic algorithm to optimize coasting points of speed-distance trajectory in order to obtain minimum energy consumption for a given travel time \cite{accikbacs2008coasting}. Yang \emph{et al.} integrated simulation-based methodologies and genetic algorithm to reduce the calculation difficulties and seek the approximate optimal coasting control strategies on the railway network \cite{yang2012optimizing}. Yin \emph{et al.} used the reinforcement learning-Q learning method, to construct an intelligent train operation system that can meet multiple objectives \cite{yin2014intelligent}.

\par Moreover, tracking the real-time train speed profile is also a critical research topic. For example, Liu \emph{et al.} proposed a high-speed railway control system based on the fuzzy control method and designed a control system in the Matlab software according to the expert experience and knowledge \cite{liu2013high}. Wu \emph{et al.} used variable structure technique and a time-delayed compensator, to design a state observer-based adaptive fuzzy controller to approximate the unknown system parameters, and thus trajectory tracking problem of a series of two-wheeled self-balancing vehicles can be addressed \cite{wu2015trajectory}. Gu \emph{et al.} proposed a new energy-efficient train operation model based on real-time traffic information from the geometric and topographic points of view through a nonlinear programming method \cite{gu2014energy}. Recently, Li \emph{et al.} designed a robust sampled-data cruise control scheduling with the form of linear matrix inequality(LMI) and proposed numerical examples that verified the effectiveness of the proposed algorithms to track the SD trajectory precisely \cite{li2014robust}.

\par Despite great achievements in previous studies, there are still some essential problems unsolved, which blocks the development of the ATO system. Firstly, for multiple objectives of train operation, most researchers mentioned above had just taken one or two objectives into account, and there is no comprehensive analysis about designing optimal train operation to meet multiple objectives. Secondly, the modern subway is capable of outputting continuous traction and braking force \cite{liu2003energy}, however, there are rare researches devoted to design the control model for continuous action while considering complicated train operation conditions, such as variables speed limits. Thirdly, in the ATO system, the optimized speed profile was designed before the operation of the train, and the train is controlled to track the designed optimized speed profile during the trip time which largely decreases the flexibility and the robustness of the ATO system. Plus, it is hard to implement complicated mathematic optimal methods to treat the nonlinear train operation problem, thus it is necessary to design a model that can realize train control without considering offline optimized speed profile. Finally, the real subway operation is faced with many unexpected situations, such as, the changed trip time of one subway which influences the timetable of the whole line, and the railway aging which changes the railway resistance condition. Being faced with these problems, modern subway always transfers from the ATO system to manual driving which largely decreases the intelligence and efficiency of train operation. From the analysis above, the contribution of the paper can be listed as follow:
\begin{itemize}
 	\item Multiple objectives of train operation are summarized and relevant evaluation indices are formulated. Through analyzing references, we summarize the expert knowledge rules and build the inference methods. They are systematically combined with reinforcement learning algorithms to help the algorithm have better performance.
 	
 	\item We establish STOD and STON based on deep deterministic policy gradient (DDPG) and normalized advantage function (NAF). On the one hand, reinforcement learning can realize model-free control. On the other hand, both DDPG and NAF are able to deal with control tasks of continuous action.
 	
 	\item The effectiveness of STOD and STON is verified by using the field data of the Yizhuang Line of the Beijing Subway (YLBS). The performance of the proposed STOD and STON is compared with the existing intelligent train operation in \cite{yin2014intelligent} and the manual driving, which illustrates that both STOD and STON have better performance than that of ITOR and manual driving. And through conducting different numerical simulations, the flexibility and the robustness of STOD and STON are proved.
 	
\end{itemize}
\par The rest of this paper is organized as follows. In Section II, we define necessary mathematic indices of train movement problems and formulate multiple objectives of train operation into numerical evaluation indices to systematically evaluate the train operation problem. Plus, we state the objectives of this paper. In Section III, the structure of STO is presented. Then we put forward expert knowledge rules and summarize inference methods. Besides, the principles and the algorithms of STOD and STON are explained. In Section IV, the simulation platform is built. Three numerical simulations are made based on the real field data of YLBS. Conclusions are summarized in Section V.

\section{Problem Formulation}
In general, the problem of train control can be formulated as an optimal control problem with focus on finding an optimal control strategy for traction and braking force during the trip time. Firstly, we define $\Delta$\(t\) as the minimum time interval, and the trip time of train can be described as follow:
\begin{equation}
t_{i+1} = t_i + \Delta t,
\end{equation}  
for \(0\leq i \leq n-1\). The total trip time \(T\) is defined as
\begin{equation}
T = t_n - t_0,
\end{equation}
where the initial running time \(t_0=0(s)\). 

\begin{figure}
	\begin{center}
		\centering
		\scriptsize
		\includegraphics[width=3.0in]{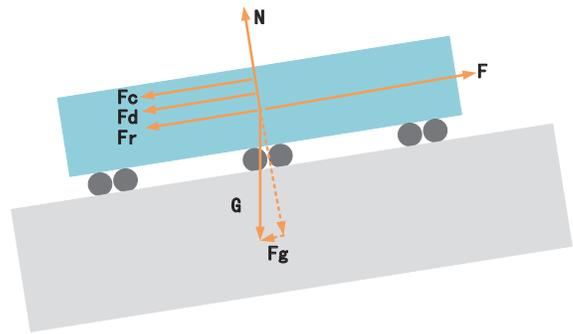}\\
	\end{center}
	\caption{The force diagram of the train.}
	\label{fig:force}
\end{figure}
\subsection{Control model of train}
The motion of the train is determined by the output force, the resistance caused by the gradient of railway, the resistance to motion, the curve resistance and the resistance caused by interactive impacts among the vehicles. According to Newton's second law, its movement equation is defined as
\begin{equation}
M(1 + \eta)u = F - F_g - F_r - F_c - F_d,
\end{equation} 
where \(M\) represents the static mass of the train; $\eta$ is the rotating factor of the train which is defined as \(\eta =M_{\eta}/M\), and \(M_{\eta}\) is the reduced mass of trains rotator; \(u\) is the acceleration or the deceleration; \(F\) is the outputted traction force or braking force of subway; \(F_g\) is the resistance caused by the gradient; \(F_r\) is  the resistance to motion given by David Equation; \(F_c\) is the curve resistance and \(F_d\) is the interactive impacts among the vehicles. The force diagram of the train is shown in Fig. \ref{fig:force}. Their definition can be described as follow:
\begin{equation}
F_g = Mgsin(\alpha(s)),
\end{equation}  
where $\alpha(s)$ represents the slope angle of the railway at position \(s\).
\begin{equation}
F_r = d_1 + d_2v + d_3 v^2,
\end{equation} 
where  \(v\) is the velocity; \(d_1\), \(d_2\) and \(d_3\) are vehicle specific coefficients which are measured by the run-down experiments\cite{rochard2000review}.
\begin{equation}
F_c = 6.3 M / [r(s) - 55],
\end{equation} 
where \(r(s)\) is the radius of the curve at the position \(s\)\cite{wang2013optimal}.
\begin{equation}
F_d = \sum_{i=1}^{k-1} (\Delta \ddot{l_i} \sum_{j=i+1}^{k} m_j),
\end{equation} 
where \(k\) is the number of vehicles; \(\Delta \ddot{l_i}\) denotes the second derivative for the distance between the center of the \(i\)th vehicle and the reference point\cite{song2011computationally}; \(m_j\) is the static mass of the \(j\)th vehicle and the static mass \(M\) of the whole train can be described as \(M=\sum_{j=1}^{k}m_j\).
\par Because there exist nonlinearity and time delay in a train control model, Eq.(8) gives the transfer function of the accelerating and decelerating process\cite{chen2013online}.
\begin{equation}
u = \frac{u_0}{1 + T_d s} e^{-T_c s},
\end{equation}
where \(u\) represents the train actual acceleration or deceleration; \(u_0\) is the accelerating or decelerating performance gain; \(T_d\) and \(T_c\) represent the time delay and the time constant of the accelerating or the decelerating process.

\subsection{The indices of model evaluation}
In general, the subway control model is generally evaluated from four aspects, i.e., the safety\cite{zheng2009modeling}, the punctuality\cite{olsson2004influencing}, the energy consumption\cite{miyatake2010optimization} and the passenger comfort \cite{karakasis2005factorial}, which are defined as follow:
\begin{figure}
	\begin{center}
		\centering
		\scriptsize
		\includegraphics[width=3.6in]{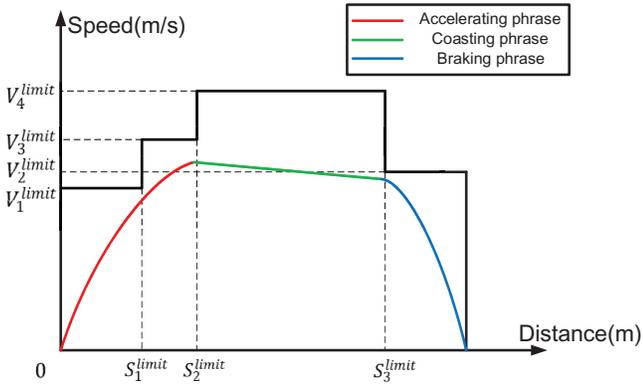}\\
	\end{center}
	\caption{Speed limits.}
	\label{fig: limit}
\end{figure}
\begin{itemize}
	\item \emph{Safety.} There may exist several speed limits between two successive subway stations and a general case is presented as in Fig. \ref{fig: limit} where \(v_1^{limit}\), \(v_2^{limit}\), \(v_3^{limit}\) and \(v_4^{limit}\) are four-speed limits of different sections between two stations. During the trip time, the velocity of the train must be inferior to the speed limit of the current section to guarantee safety. The safety evaluation index \(I_s\) is defined as 
    \begin{equation}
    I_s=\left\{
    \begin{array}{rcl}
    1 & & v_i \leq v_i^{limit}(\forall i)\\
    0 & & v_i > v_i^{limit}(\exists i) \\
    \end{array} 
    \right..
    \end{equation}
	\item \emph{Punctuality.} Punctuality is a very important factor in train operation. As the time interval between two adjacent subways is very short, the accidental delay problem may influence the timetable of the whole line. We first define the running time error \(e_{t}\) as
	\begin{equation}
	e_{t}={|T_{actual}-T_{planning}|},
	\end{equation}	
	where \(T_{actual}\) is the actual running time of the train, and \(T_{planning}\) is the planning trip time of the train. In this paper, if the running time error \(e_{t}\) is superior to \(3s\), the subway is not punctual. Thus the definition of the punctuality evaluation index \(I_t\) is given as
    \begin{equation}
	I_t=\left\{
	\begin{array}{rcl}
	1 & & e_{t} \leq 3\\
	0 & & e_{t} > 3\\
	\end{array} 
	\right..
	\end{equation}
	\item \emph{Energy efficiency.} The energy efficiency is one of the focuses of modern society and the energy consumption makes up a large part of the cost of train operation. These concerns make energy efficiency play a core actor in our control model designing. According to \cite{cheng2017intelligent}, the equation to calculate the consumed energy \(E\) is described as 
	\begin{equation}
	E=\sum_{i=1}^n( M |u_i| v(t_i) \Delta t).
	\end{equation}
	Based on the equation of consumed energy, we define the energy efficiency evaluation index \(I_e\) as 
	\begin{equation}
	I_e= \frac{E}{M}. 
	\end{equation}
	\item \emph{Riding comfort.} The riding comfort is a direct evaluation criterion for train service quality \cite{powell2015passenger}, and it guarantees that the instantaneous change of acceleration or deceleration should below a certain threshold. We define the jerk or the rate of change of acceleration $\Delta u$ as:
	\begin{equation}
	\Delta u_{i} = |\frac{u_{i}-u_{i-1}}{\Delta t}|.
	\end{equation}
	Thus the riding comfort evaluation index \(I_c\) can be defined as:
	\begin{equation}
	I_c=\sum_{i=1}^n\left\{
	\begin{array}{rcl}
	0 & & \Delta u_{i} \leq \Delta U'\\
	\Delta u_{i} & & \Delta u_{i} > \Delta U'\\
	\end{array} 
	\right.,
	\end{equation}
	where $\Delta U'$ is the threshold for change of acceleration, in this case, $\Delta U' = 0.30 m/s^3$ as proposed in \cite{hoberock1977survey}.
\end{itemize}

\subsection{Problem statement}
Two designed STO algorithms, including STOD and STON, are supposed to achieve four purposes. Firstly, STO algorithms can provide the control strategy for the traction force and the braking force which can meet the basic requirements, including the safety and the punctuality for train operation. Secondly, STO algorithms can perform properly without considering offline designed speed profile and realize the control for the continuous force. Thirdly, the control strategy outputted by STO algorithms can outperform experienced subway drivers in the aspect of energy efficiency while ensuring good riding comfort. Last, STO algorithms can adapt to different situations including different trip times and different resistance conditions.    
\par The fact that the existed ATO system has to track the designed offline speed profile and modern subway can output continuous traction and braking force, have motivated our study. Moreover, reinforcement learning has been applied to many fields to deal with the model-free problem \cite{ruelens2015learning} and expert knowledge has also been largely used to improve control strategy \cite{cheng2017intelligent}. Hence, we have put forward two STO algorithms based on the fusion of expert knowledge and reinforcement learning.

\section{Design of intelligent train control model}
In this section, we will give a detailed explanation for STO algorithms, including the structure of STO, the expert knowledge rules, the inference methods, and the principles for STOD and STON.
\subsection{The structure of intelligent control model}
\begin{figure*}
	\begin{center}
		\centering
		\scriptsize
		\includegraphics[width=6.0in]{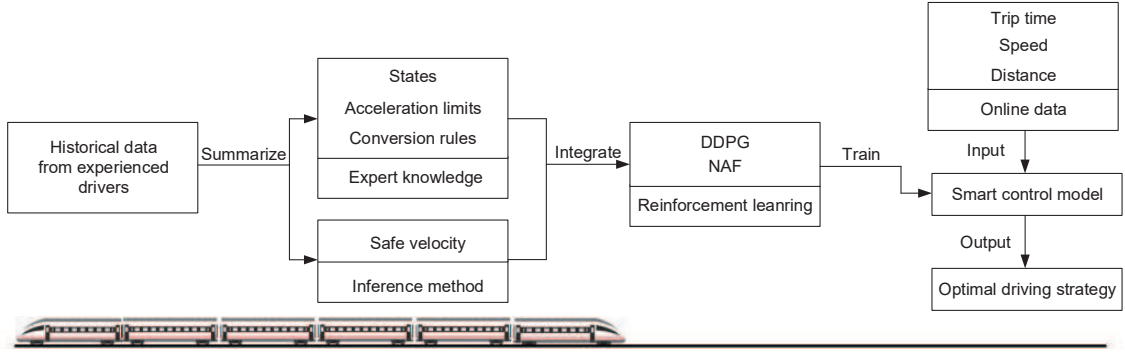}\\
	\end{center}
	\caption{STO model.}
	\label{fig: STO}
\end{figure*}

The structure of STO is shown in Fig. \ref{fig: STO}. We can learn that the STO model contains three phases. The first phase is to obtain expert knowledge and to develop inference methods which are essential for building a stable model. The second phase is to integrate expert knowledge and heuristic inference methods into the reinforcement algorithms. The third phase is to train designed algorithms and get a stable model.

\par For the real-world application of STO, it usually includes the last two steps. Above all, we will follow three phases of STO to establish a stable model before putting it into practice. Then, during the trip time of the train equipped with STO, the system will accept the real-time information about its position, its velocity and its running time obtained from onboard sensors, and will send the command about the control of traction or braking force.

\subsection{Expert knowledge and heuristic inference rules}
As the train control problem has features of non-linearity and complexity, it is hard to design an ideal model without taking expert knowledge into account. And knowledge-based technology has been successfully applied to solve complicated optimal control problem\cite{murrell1997survey}. According to the previous works \cite{yin2016smart}\cite{cheng2017intelligent}\cite{zhao2017integrated}, we found that those optimal train operation methods always follow certain expert rules which can be listed as: 
\begin{itemize}
	\item The train operation has three states, including the accelerating, the coasting, and the braking as shown in Fig. \ref{fig: limit}. Unless encountering special accidents, the train wouldn't transfer directly from accelerating state to braking state and vice verse. The transfer between any other two states is allowable.   
	\item For the sake of protecting the engine and ensuring the riding comfort, the acceleration of the train shouldn't exceed \(0.6 m/s^2\) when the subway starts its operation. 
\end{itemize}	

\par Besides, experienced drivers know well about when the train should decelerate to guarantee the safety of train operation, and there is no designed speed profile for STO. Thus, we develop the heuristic inference method to ensure that the designed model will work properly. The heuristic inference method is listed as:

\begin{itemize}
	\item When the velocity limit \(v_{j+1}^{limit}\) of the next section \(j+1\) is less than the velocity limit \(v_{j}^{limit}\) of the current section \(j\) shown in Fig. 3, the train may have to rationally brake to guarantee the safety of the train. In other words, the speed of the train should always be inferior to the speed limit. In this case, we define the safe velocity \(v_{i}^{safe}\) to supervise the speed of the train as
	\begin{equation}
	v_{i}^{safe}=\sqrt{\beta(v_{j+1}^{limit})^2-2  u_{min}(s_{j+1}^{limit}-s_i)},
	\end{equation}
    where \(s_i\) is the current position ot the train. \(s_{limit(j+1)}\) is the starting position of the next section. \(\beta\) is a speed proportional coefficient caused by the time delay and the friction of the railway \cite{yin2016smart}. $u_{min}$ is the minimum deceleration. In this case, $u_{min} = -1m/s^2$. If the current velocity \(v_i\) is superior or equal to \(v_s(i)\), the train should adapt minimum deceleration immediately.
\end{itemize}	

\subsection{Continuous action control methods based on reinforcement learning}
As the expert knowledge cannot allow the agent to perform better than experienced drivers and it does not have the learning process, we combine the expert knowledge with the reinforcement learning methods. In this way, it can ensure that outputted control strategy meets basic requirements and it can also have the possibility to find an optimal solution. Moreover, as two popular reinforcement learning algorithms for the control of continuous action, DDPG and NAF have their advantages in different fields. To compare their performance, in this work, we have designed STOD and STON. 
\par Reinforcement learning can allow agents to automatically take proper action by maximizing their reward\cite{sutton1998introduction}. As a powerful decision-making tool\cite{kaelbling1996reinforcement}, reinforcement learning has been used to deal with optimal control problems in many different fields, such as aerobatic helicopter flight control \cite{abbeel2007application}, playing robot soccer game\cite{duan2007application}, power systems stability control\cite{ernst2004power} and so on. There are two reasons which drive us to adapt reinforcement learning in the train control task. Firstly, some reinforcement learning algorithms can realize the control for continuous action\cite{lillicrap2015continuous} which can improve the current control strategy for discrete action in the ATO system. Secondly, reinforcement learning pays attention to long-term rewards, while the train's current action also influences its follow-up steps.
\begin{figure}
	\begin{center}
		\centering
		\scriptsize
		\includegraphics[width=3.6in]{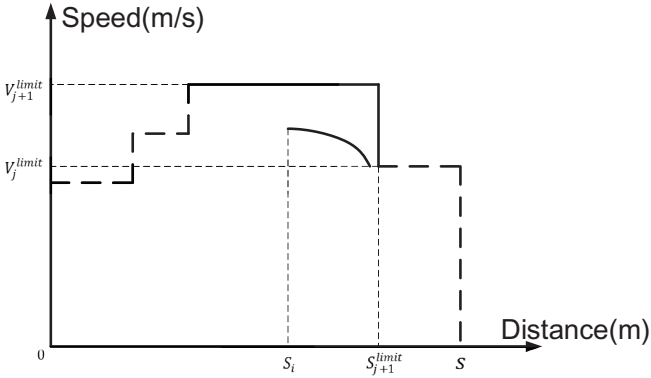}\\
	\end{center}
	\caption{Safety velocity.}
\end{figure}

\subsubsection{Markov decision process}
Before applying the reinforcement learning algorithm, we formulate our problem into a Markov decision process (MDP) which provides a mathematical framework for decision making. The critical elements of reinforcement learning include its state, action, policy, and reward, which are defined as follows:
\begin{itemize}
	\item \emph{State \(x\).} In this case, the speed and the velocity, two important train movement factors, make up the state. Thus, it can be described as
	\begin{equation}
	x_i = [s_i, v_i],	
	\end{equation}
	where \(0 \leq i \leq n\). And the initial state \(x_0\) is defined as
	\begin{equation}
	x_0 = [0, 0].	
	\end{equation}	
	\item \emph{Action \(a\).} During the trip time, the acceleration \( u_i\) is defined as the action. And the range of acceleration is defined as: $u_i \in [-1,1]$ for the subway operation in YLBS. Thus the Action \(a\) is defined as
	\begin{equation}
	a_i = [u_i].	
	\end{equation}	
	\item \emph{Policy \(\pi\).} The policy \(\pi\) denotes the probability of taking an action when dealing with a discrete action task. In this paper, as STO is designed to deal with continuous action control task, the policy \(\pi\) is the statistics of the probability distribution. It can be expressed as
	\begin{equation}
	\pi(a|x,\theta) = N(\mu(x,\theta),sigma(x,\theta)),
	\end{equation}
	where \(\theta\) is the weight.
	\item \emph{Reward function \(r(x_i, a_i)\).} This function defines the reward obtained by the train when it takes an action at a certain state. In this case, our reward function is defined by the energy consumption per weight \(\Delta I_e\) during the time interval \(\Delta t\) when train takes the action \(a_i\) at the state \(x_i\), and the time error \(e'_{ti}\). The time error \(e'_{ti}\) is used to ensure that the agent should arrive at the destination within planning trip time rather than spending too much on running with low speed to minimize the energy consumption. The reward function is defined in
	\begin{equation}
	r(x_i, a_i) =  -\lambda_1 \Delta I_e - \lambda_2 e'_{ti} - \lambda_3 \Delta I_c - \lambda_4 D - \lambda_5 Acc,
	\end{equation}	
	where $\lambda_1$, $\lambda_2$, $\lambda_3$, $\lambda_4$ and $\lambda_5$ are the coefficients defined to meet different requirements of system; the time error \(e'_{ti}\) at moment \(t_i\) is defined as
	\begin{equation}
    e'_{ti}=\left\{
    \begin{array}{rcl}
    1 & & t_i > T_{planning}\\
    0 & & t_i \leq T_{planning}\\
    \end{array} 
    \right.;
    \end{equation}
    $D$ is used to check whether the train has arrived the destination and stopped at the correct position. Its expression can be written as
    \begin{equation}
    D = \left\{
    \begin{array}{rcl}
    1 & & s_i > S_{Destination}\\
    0 & & s_i \leq S_{Destination}\\
    \end{array} 
    \right.;
    \end{equation}
    $Acc$ is used to guarantee that the running time in the expected range and it can be defined as
    \begin{equation}
    Acc = D * |t_i - T_{planning}|.
    \end{equation}
    Noticing $D$ equals to $1$ only when the train arrived, thus this equation can use to calculate the difference between the whole running time and the planning trip time.
\end{itemize}
\subsubsection{STOD} 
STOD algorithm is based on the reinforcement learning algorithm DDPG which is an actor-critic, model-free algorithm that can deal with continuous action control problems, based on policy-gradient algorithm\cite{kaelbling1996reinforcement}. The reinforcement learning setup consists of an agent interacting with an environment \(E\) and we denote the discounted state visitation distribution for a policy \(\pi\) as \(\rho ^{\pi} \). In the DDPG, the critic-network is used to estimate the action-value function, while the actor-network is used to improve the policy function with the help of critic-network. Besides, we use \(\theta^{Q}\) to represent the weight of action-value function \(Q(x,a|\theta^{Q})\) and use \(\theta^{\mu}\) to represent the weight of policy function \(a=\mu (x|\theta^{\mu})\). The loss function \(L\) for critic-network is described in Eq.(22), and \(\theta^Q\) is updated through minimizing the loss function.
\begin{equation}
L(\theta) = \mathbb{E}_{x_i \sim \rho ^\beta, a_i \sim \beta, r_i \sim E}[(Q(x_i,a_i|\theta^Q)-y_i)^2],
\end{equation}
where \(\beta\) represents a stochastic behavior policy, and target value \(y_i\) is described as
\begin{equation}
y_i=r(x_i,a_i)+\gamma Q(x_{i+1},\mu(x_{i+1})|\theta ^ Q),
\end{equation}
where \(\gamma\) describes the discount rate. 
\par  The return from a state \(x_i\) is defined as the sum of the future discounted reward in
\begin{equation}
R_i =  \sum_{j=i}^n \gamma^{j-i} r(x_j,a_j).
\end{equation}
\par The goal of actor-network is to maximize the return from the start distribution \(J=\mathbb{E}_{r_i,x_i\sim E,a_i\sim\pi}[R_1]\). In the traditional Q-learning, the network $Q(x,a|\theta^Q)$ is used to calculate target value $y_i$ and is also updated based on the target value. This method will increase the instability of the Q network, as during the training the process, the Q network is constantly updated. If we use a constantly changing value as our target value to update the network, the feedback loops between the target and estimated Q-values will destabilize the Q network \cite{mnih2013playing} \cite{mnih2015human}. To solve this problem, the target network is implemented. In DDPG, there is a target actor-network \(\mu'(x|\theta^{\mu'})\) and a target critic network \(Q'(x,a|\theta^{Q'})\). Their weight \(\theta^{\mu'}\) and \(\theta^{Q'}\) are updated by the following equations:
\begin{equation}
\theta^{\mu'} \leftarrow \tau \theta^{\mu} + (1-\tau)\theta^{\mu'}
\end{equation}
and
\begin{equation}
\theta^{Q'} \leftarrow \tau \theta^{Q} + (1-\tau)\theta^{Q'},
\end{equation}
where \(\tau \ll 1\). It indicates that the weights of two target networks are updated more slowly than the weights of the actor-network and the critic network, which can improve the stability of the learning process. The STOD algorithm is given in Algorithm 1.

\begin{algorithm}
	\caption{\quad STOD algorithm}
	// Initilize parameters of STOD
	\par Randomly initialize normalized actor network \(\mu(x|\theta^{\mu})\) and initialize target actor network \(\theta^{\mu'} \leftarrow \theta^{\mu} \)
	\par Randomly initialize normalized critic network \(Q(x,a|\theta^{Q})\) and initialize target critic network \(\theta^{Q'} \leftarrow \theta^{Q} \)
	\par Initialize reply buffer \(R \leftarrow \emptyset\)
	\begin{algorithmic}
    \STATE // Excute the networks 
	\FOR{episode=\(1, M\) do} 
    \STATE  Initialize a random process \(\mathcal{N}\) for action exploration 
	\STATE  Initialize observation state \(x_0 \leftarrow [0,0]\) 
    \FOR{\(i\)=\(1, N\) do}   
    \STATE Obtain action \(a_i=\mu(x_i|\theta^{\mu}) + \mathcal{N}_i\)
    \STATE Verify the obtained action \(a_i\) with expert knowledge and inference method. If action \(a_i\) doesn't meet those requirements, adjusting the obtained action \(a_i\) 
    \STATE Execute action \(a_i\) and observe reward \(r_i\) and state \(x_{i+1}\)according to the subway motion equation 
    \STATE Store transition \((x_i,a_i,r_i,x_{i+1})\) in buffer \(R\)
    \STATE // Update the weights 
    \STATE  Randomly sample a minibatch of \(N\) transitions \((x_j,a_j,r_j,x_{j+1})\) from buffer \(R\)
    \STATE Calculate:
    \(y_j=r_j + \gamma Q'(x_{j+1},\mu'(x_{j+1}|\theta^{\mu'})|\theta^{Q'})\)
    \STATE Update critic network by minimizing the loss function:
    \STATE
    \(L =\frac{1}{N}\sum_{j}(Q(x_j,a_j|\theta^Q)-y_j)^2\)
    \STATE Update actor network through policy gradient:
    \STATE
    \(\nabla_{\theta^\mu} J\approx \frac{1}{N} \sum_{j} \nabla_aQ(x,a|\theta^Q)|_{x=x_j,a=\mu(x_j)} 
    \nabla_{\theta^\mu}\mu(x|\theta^{\mu})|s_j \)
    \STATE Update the target networks:
    \STATE
    \(\theta^{Q'} \leftarrow \tau \theta^{Q} + (1-\tau)\theta^{Q'}\)
    \STATE
    \(\theta^{\mu'} \leftarrow \tau \theta^{\mu} + (1-\tau)\theta^{\mu'}\)
    \ENDFOR
    \ENDFOR
    \end{algorithmic}
\end{algorithm} 
\subsubsection{STON}
STON algorithm is based on the reinforcement learning algorithm NAF which is a reinforcement learning method designed for continuous control tasks and works as an alternative to commonly used policy gradient and actor-critic methods, such as DDPG. Plus, it allows users to use the Q-learning method to deal with the control tasks for continuous action, thus STON algorithm is simpler than the STOD algorithm. Q-learning is not suitable for dealing with continuous action tasks, as it should maximize a complex, nonlinear function at each update. And the idea behind NAF is to represent the Q-function \(Q(x_i,a_i)\) in the way that its maximum, \(argmax_a Q(x_i, a_i)\) can be determined during the Q-learning update \cite{gu2016continuous}. In NAF, the neural network output separately the value function \(V(x)\) and the advantage term \(A(x,a)\), which are defined as follow:
\begin{equation}
Q(x,a|\theta^Q) = A(x,a|\theta^A) + V(x|\theta^v),
\end{equation}
and
\begin{equation}
A(x,a|\theta^A) = - \frac{1}{2} (a-\mu(x|\theta^{\mu}))^T P(x|\theta^{p})(a-\mu(x|\theta^{\mu})).
\end{equation}
\par \(P(x|\theta^P)\) is a state-dependent, positive-definite square matrix. With the Cholesky decomposition method, it can be described as
\begin{equation}
P(x|\theta^P) = L(x|\theta^P)L(x|\theta^P)^T,
\end{equation}
where \(L(x|\theta^P)\) is a lower-triangular matrix outputted by the neural network. And the network is updated by minimizing its loss function \(L=\frac{1}{N}\sum_{i}(y_i - Q(x_i, a_i|\theta^Q))^2\). In this algorithm, the target network will also be introduced to improve the stability of the learning process. And the STON algorithm is described in algorithm 2.
\begin{algorithm}
	\caption{\quad STON algorithm}
	// Initilize parameters of STON
	\par Randomly initialize normalized \(Q\) network \(Q(x,a|\theta^{Q})\) and initialize target \(Q'\) network \(\theta^{Q'} \leftarrow \theta^{Q} \)
	\par Initialize reply buffer \(R \leftarrow \emptyset\)
    \begin{algorithmic}
	\STATE // Excute the networks 
	\FOR{episode=\(1, M\) do}
	\STATE  Initialize a random process \(\mathcal{N}\) for action exploration 
	\STATE  Initialize observation state \(x_0 \leftarrow [0,0]\) 
	\FOR{\(i\)=\(1, N\) do}    
	\STATE Obtain action \(a_i=\mu(x_i|\theta^{\mu}) + \mathcal{N_i}\)
    \STATE Verify the obtained action \(a_i\) with expert knowledge and inference method. If action \(a_i\) doesn't meet those requirements, adjusting the obtained action \(a_i\) 
	\STATE Execute action \(a_i\) and observe reward \(r_i\) and state \(x_{i+1}\) according to the subway motion equation 
	\STATE Store transition \((x_i,a_i,r_i,x_{i+1})\) in buffer \(R\)
	\STATE // Update the weights
	\FOR{iteration = 1, I do}
	\STATE  Randomly sample a minibatch of \(m\) transition \((x_j,a_j,r_j,x_{j+1})\) from buffer \(R\)
	\STATE Calculate:
	\(y_j=r_j + \gamma V'(x_{j+1}|\theta^{Q'})\)
	\STATE Update critic network by minimizing the loss function:
	\STATE
	\(L =\frac{1}{N}\sum_{j}(y_j - Q(x_j,a_j|\theta^Q))^2\)
	\STATE Update the target networks:
	\STATE
	\(\theta^{Q'} \leftarrow \tau \theta^{Q} + (1-\tau)\theta^{Q'}\)
	\ENDFOR
    \ENDFOR
	\ENDFOR
	\end{algorithmic}
\end{algorithm} 
\subsubsection{ITOR}
ITOR algorithm is a new ATO algorithm proposed in \cite{yin2014intelligent}, which employs the deep Q-learning algorithm to construct the framework. Due to the limitation of deep Q-learning, ITOR can only realize the discrete action control of the train, whose performance will be compared with STOD and STON in the latter experiment.
\section{Simulations}
\begin{figure}
	\begin{center}
		\centering
		\scriptsize
		\includegraphics[width=3.0in]{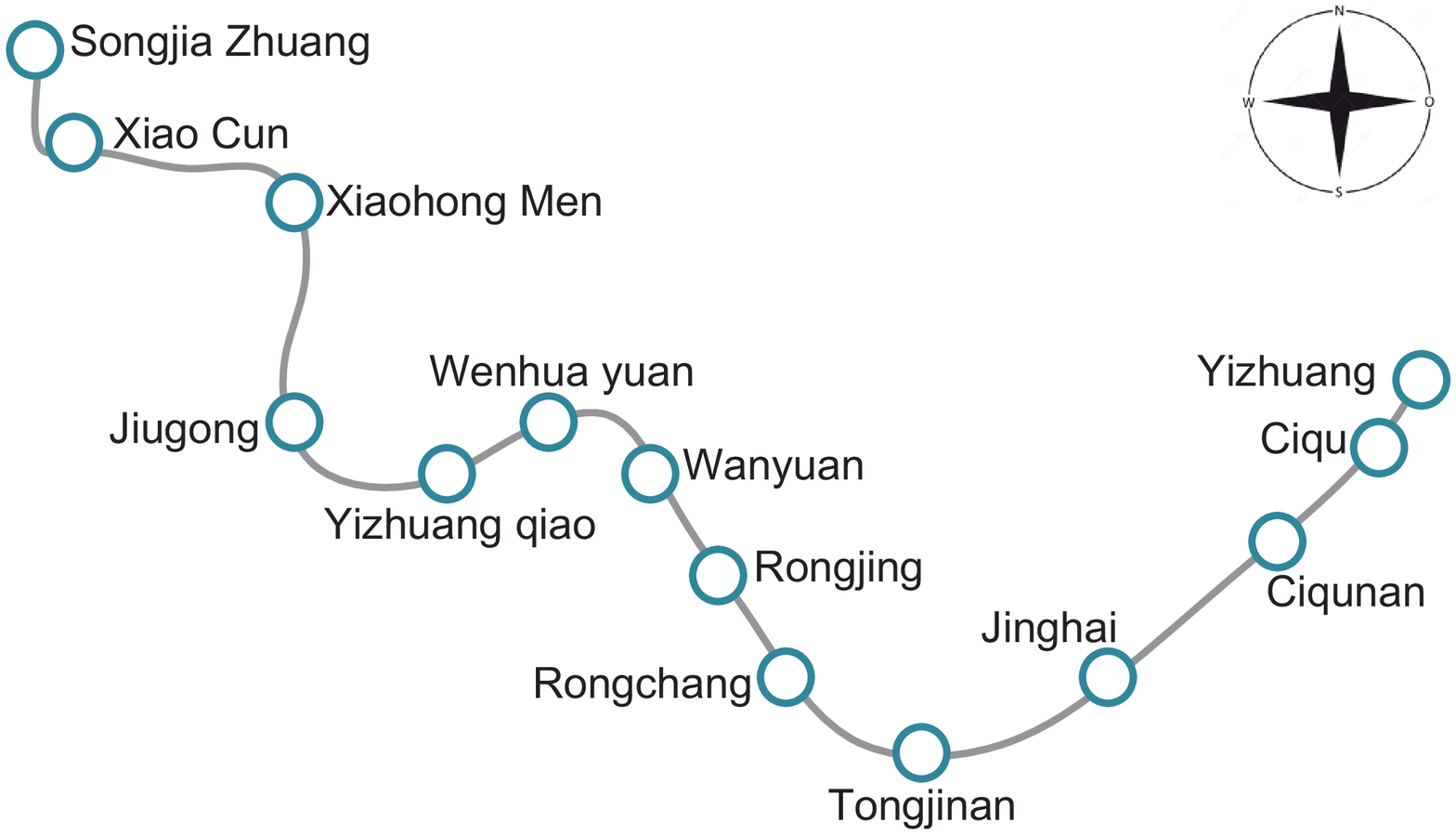}\\
	\end{center}
	\caption{YBLS.}
	\label{fig: YBLS}
\end{figure}

To verify the effectiveness, the flexibility and the robustness of STOD and STON, we have designed three numerical simulation experiments based on real field data collected from YLBS. The YLBS started operation on 30th December 2010 in Beijing. The total length of YLBS is up to \(23.3 km\) and it starts from Songjiazhuang station and ends at Ciqu station as in Fig. \ref{fig: YBLS}. The type of train used in YLBS is DKZ32 EMU which has six vehicles and its parameters are given in Table. \ref{tab: tab1}. 

\par During the training process of ITOR, STOD, and STON, we employ the Adam optimizer and set the learning rate as $1 \times 10^{-4}$ for all the networks training, except the training for the critic network of the STOD whose learning rate is $5 \times 10^{-5}$. The $\tau$ in the target network is set as $1 \times 10^{-3}$. The discount factor for the value function is $0.99$ and the size of the mini-batch for the memory reply is $256$. As to the weight coefficients of the reward function, $\lambda_1 = 0.13$, $\lambda_2 = 30$, $\lambda_3 = 10$, $\lambda_4 = 400$ and $\lambda_5 = 70$.

For the ITOR, it has five hidden layers. The first hidden layer has 400 units; the second layer has 300 units; the third layer has 200 units; the fourth layer has 100 units and the last layer has 32 units. Each one of the first four hidden layers is followed by a Relu activation function and the last hidden layer doesn't have any activation layer. For parameters of STOD, both its actor-network and critic network have five hidden layers. The first layer has 400 units; the second layer has 300 units; the third layer has 200 units; the fourth layer has 100 units and the last layer has 32 units. Each one of the first four hidden layers is followed by a Relu activation function; the last hidden layer of actor-network is followed by a Tanh activation function; the last hidden layer of critic network doesn't have any activation layer. The target actor-network shares the same structure with the actor-network, and the target critic network shares the same structure with the critic network. For the STON, it has five hidden layers. The first hidden layer has 400 units; the second layer has 300 units; the third layer has 200 units; the fourth layer has 100 units and the last layer has 32 units. Each one of the first four hidden layers is followed by a Tanh activation function. The target network shares the same structure.

\par In this section, we will present the simulation result of three cases. In \emph{case 1}, a comparison between manual driving data of the experienced driver, ITOR, STOD, and STON, is derived. In \emph{case 2}, we test the flexibility of ITOR, STOD, and STON by changing the planning trip time of the same railway section. In \emph{case 3}, we alter the gradient condition of the railway section to verify the robustness of ITOR, STOD, and STON.

\begin{table}[htbp]
	\caption{Parameters of DKZ32}
	\begin{center}
		\begin{tabular}{c|c}
			\hline
			Parameters & Value \\
			\hline
			M (kg) & \(1.99 \times 10^5\) \\
			\(m_i, i =1,6\) (kg) & \(3.3 \times 10^4\) \\
            \(m_i, i=3\) (kg) & \(2.8 \times 10^4\) \\
            \(m_i, i=2,4,5\) (kg) & \(3.5 \times 10^4\) \\
			\(\Delta l_i,i=1,3,6\)(mm) & \(0.1sin(t)\) \\
			\(\Delta l_i,i=2,4,5\)(mm) & \(0.15cos(t)\) \\
			Time constant (Braking) \(T'_c\) & \(0.4\)  \\
			Time delay (Braking) \(T'_d\) & \(0.8\)  \\
			Time constant (Accelerating) \(T_c\) & \(0.4\)  \\
			Time delay (Accelerating) \(T_d\) & \(1\)  \\
			\(d_1, d_2, d_3\)  & \(1.244, 1.45 \times 10^{-2},1.36 \times 10^{-4} \) \\
			\hline
		\end{tabular}
		\label{tab1}
	\end{center}
	\label{tab: tab1}
\end{table}
\subsection{Case 1}
\begin{figure}
	\begin{center}
		\centering
		\scriptsize
		\includegraphics[width=3.6in]{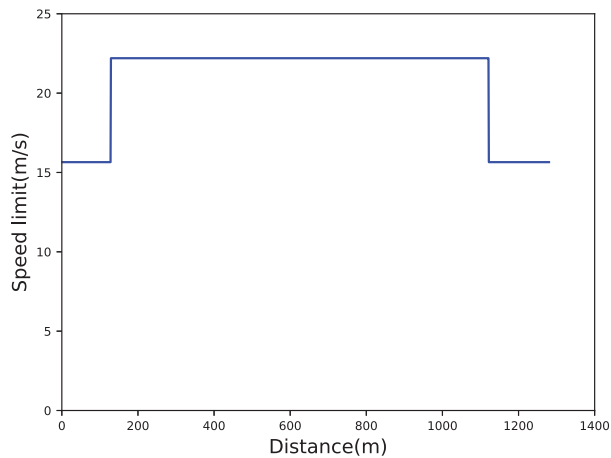}\\
	\end{center}
	\caption{Speed limits between Rongjing East Street station and Wanyuan Street station.}
	\label{fig: limit2}
\end{figure}

\begin{figure}
	\begin{center}
		\centering
		\scriptsize
		\includegraphics[width=3.6in]{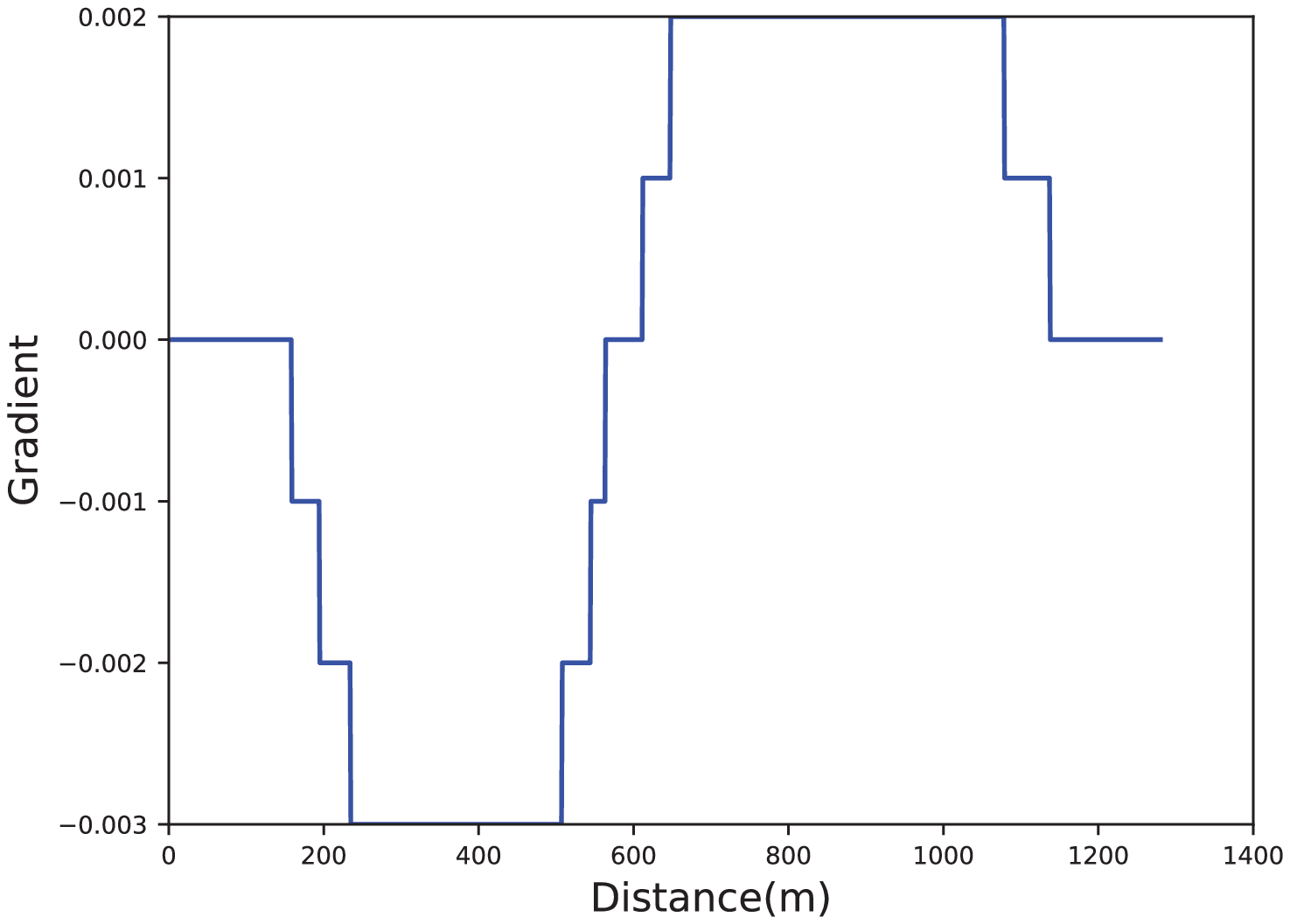}\\
	\end{center}
	\caption{Gradient condition between Rongjing East Street station and Wanyuan Street station.}
	\label{fig: GR}
\end{figure}

In the first case, we use the field data collected railway section between Rongjing East Street station and WanyuanStreet station in YLBS. The whole length of this section is 1280m and the planning trip time is 101s. The speed limit information of this section is shown in Fig. \ref{fig: limit2} and the gradient condition of this section are described in Fig. \ref{fig: GR}. In order to find the best manual driving data, we have analyzed 100 groups of up trains and down trains of this section from May 1, 2015, to May 27, 2015.

\par
\begin{figure}
	\begin{center}
		\centering
		\scriptsize
		\includegraphics[width=3.6in]{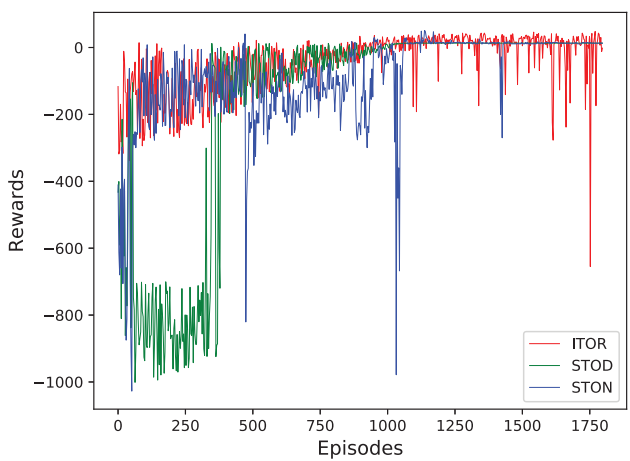}\\
	\end{center}
	\caption{Learning curve of the ITOR, STOD and STON.}
	\label{fig: learning}
\end{figure}

\begin{figure}
	\begin{center}
		\centering
		\scriptsize
		\includegraphics[width=3.6in]{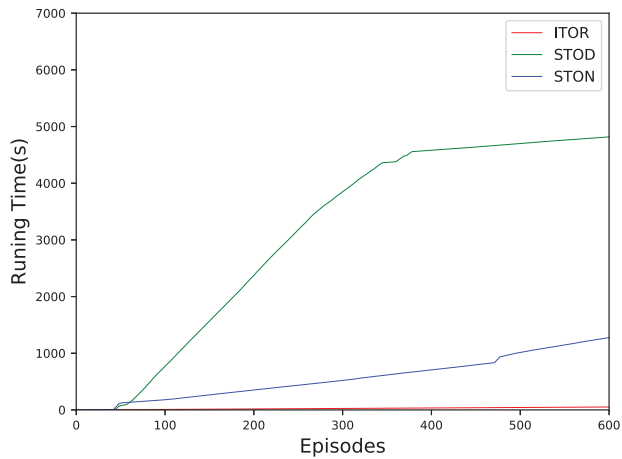}\\
	\end{center}
	\caption{Runing time for the learning process of the ITOR, STOD and STON.}
	\label{fig: time}
\end{figure}

 Fig. \ref{fig: learning} shows the training process for ITOR, STOD, and STON. Due to our definition of our reward, the reward always begins with a negative number. From Fig. \ref{fig: learning}, all three algorithms reach a relatively stable phase after $1050$ episodes. However, one thing worth to be noticed is that compared with ITOR, STOD and STON always perform better with a higher reward. Even after 1750 episodes, ITOR algorithm still fluctuates a lot. Thus STOD and STON are more stable than ITOR. Fig. \ref{fig: time} shows the running time of the three algorithms. Within our expectation, we find that ITOR has less running time due to its simple structure and its discrete action space. STOD and STON have longer running time, because of their more complicated network structure and their continuous action space. 

\par
\begin{figure}
	\begin{center}
		\centering
		\scriptsize
		\includegraphics[width=3.6in]{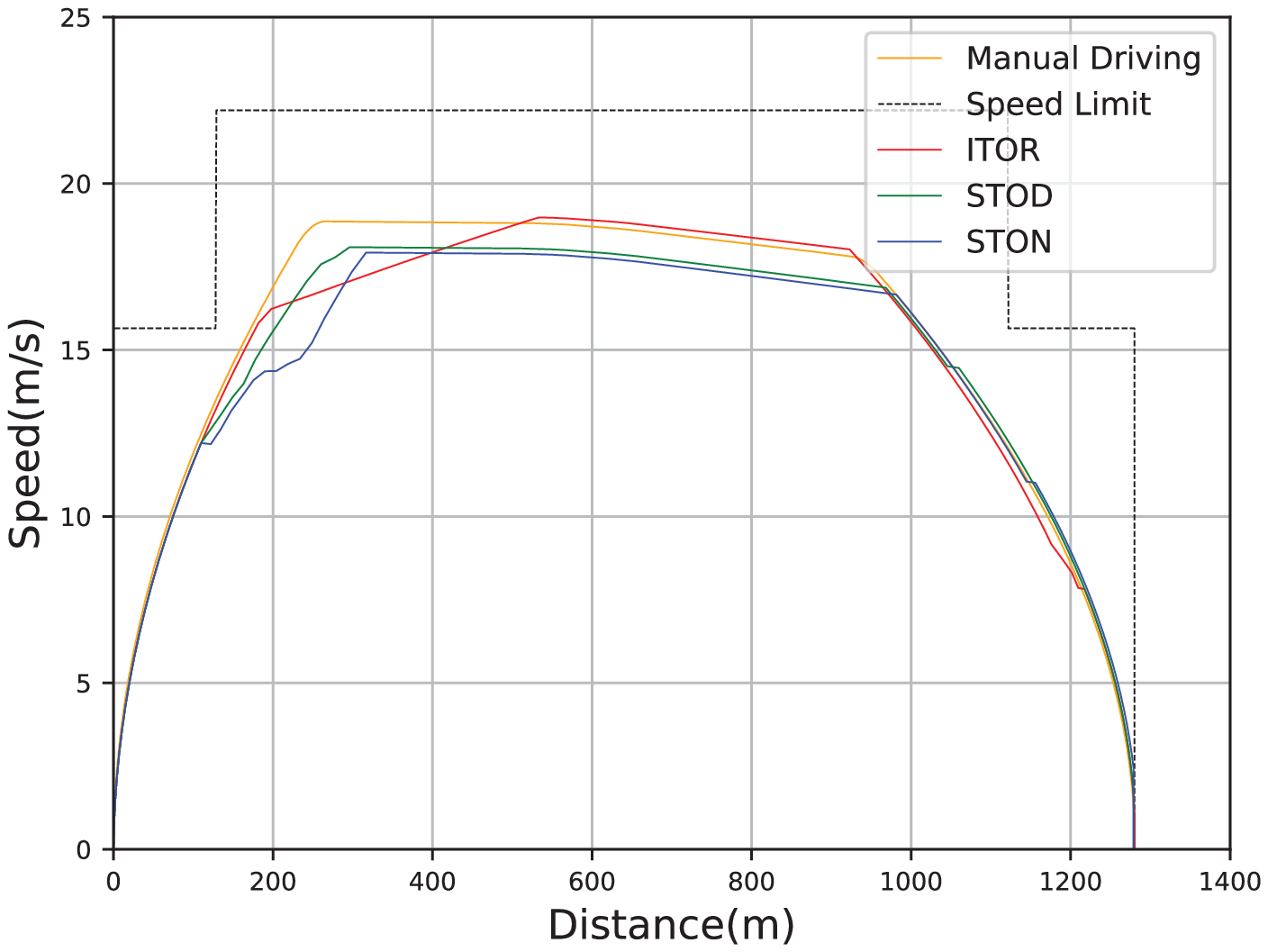}\\
	\end{center}
	\caption{Comparison of speed distance profile with \(101s\) trip time.}
	\label{fig: 101}
\end{figure}

 
Fig. \ref{fig: 101} presents speed distance profiles for $101s$ trip time of the four models. We can learn that the speed profile of the manual driving can be obviously divided into a full accelerating phase, a coasting phase, and a full braking phase. As to the maximum speed, manually driving speed profile has the maximum speed $18.86m/s$. The speed-distance profile of ITOR has the highest maximum speed compared with other three profiles. Its profile can be divided into four phases including a full accelerating phrase, an accelerating phase, a coasting phrase, and a full braking phrase. Its maximum speed is $18.98m/s$. The speed-distance profiles of STOD and STON are similar. Both of them have a lower maximum speed than that of manual driving and ITOR, which indicates that they may have lower energy consumption. The maximum speed of STOD is $18.08m/s$, while the maximum speed of STON is $17.93m/s$. 

\par Table. \ref{tab2} gives the comparison in the aspect of four evaluation indices and the running time. We can learn that the punctuality evaluation indices of four frameworks satisfy the requirement of YLBS.  As to the safety evaluation indices, all four models meet the requirement. Among the four models, ITOR has the largest energy consumption. Compared with the manual driving, ITOR costs $1.7\%$ of energy more than the manual driving; STOD costs $9.4\%$ of energy less than manual driving and STON costs $11.7\%$ of energy consumption less than manual driving. In the aspect of the riding comfort for four models, the manual driving has the highest $I_c$ which indicates the worst passenger comfort, while ITOR, STOD, and STON have similar value for riding comfort evaluation index which is much smaller than that of the manually driving. Among them, STON has the best $I_e$ and $I_c$ with the trip time $101s$.
 
\begin{table}[htbp]
	\caption{Comparison of performance with different trip time and gradient condition}
	\begin{center}
		\begin{tabular}{c|c|c|c|c|c}
			\hline
			Evaluation Indices & $t$ & $I_t$ & $I_s$ & $I_e$ & $I_c$ \\
			\hline
            Manual Driving (101s) & 102s & 1 & 1 & 586.82 & 9.05 \\
            ITOR Driving (101s) & 102s & 1 & 1 & 597.21 & 4.60 \\
            STOD (101s) & 102s & 1 & 1 & 531.77 & 4.58 \\ 
            STON (101s) & 102s & 1 & 1 & \textbf{518.03} & \textbf{4.56} \\  
			\hline
			\hline
            Manual Driving (95s) & 96s & 1 & 1 & 811.77 & 14.00 \\
            ITOR Driving (95s) & 97s & 1 & 1 & 854.24 & 8.8 \\
            STOD (95s) & 96s & 1 & 1 & 741.01 & 5.84 \\
            STON (95s) & 96s & 1 & 1 & \textbf{740.29} & \textbf{5.27} \\
			\hline
			\hline
			Manual Driving (115s) & 116s & 1 & 1 & 325.05 & 7.50 \\
            ITOR Driving (115s) & 116s & 1 & 1 & 326.58 & \textbf{3.80} \\
            STOD (115s) & 116s & 1 & 1 & 344.76 & 5.80 \\ 
            STON (115s) & 115s & 1 & 1 & \textbf{320.06} & 4.01 \\  
			\hline
			\hline
			ITOR (New Gradient) & 102s & 1 & 1 & 619.11 & 4.60 \\
			STOD (New Gradient) & 102s & 1 & 1 & 568.56 & 4.41 \\
			STON (New Gradient) & 103s & 1 & 1 & \textbf{467.62} & \textbf{2.23} \\ 
			\hline
		\end{tabular}
		\label{tab2}
	\end{center}
\end{table}

\subsection{Case 2}
We verify the flexibility of ITOR, STON, and STOD through conducting the simulation with the different trip times in the same railway section. As in the real-time subway operation, the technical accident and the large crowd during the morning and evening can largely influence the trip time of the subway. To ensure the normal operation of the whole line, the subway needs to change its control strategy. We conducted two simulations with different trip times, including one with $95s$ planning trip time and one with $115s$ planning trip time. In this subsection, we will compare the performance of manually driving, ITOR, STOD, and STON with different planning trip times.

\begin{figure}
	\begin{center}
		\centering
		\scriptsize
		\includegraphics[width=3.6in]{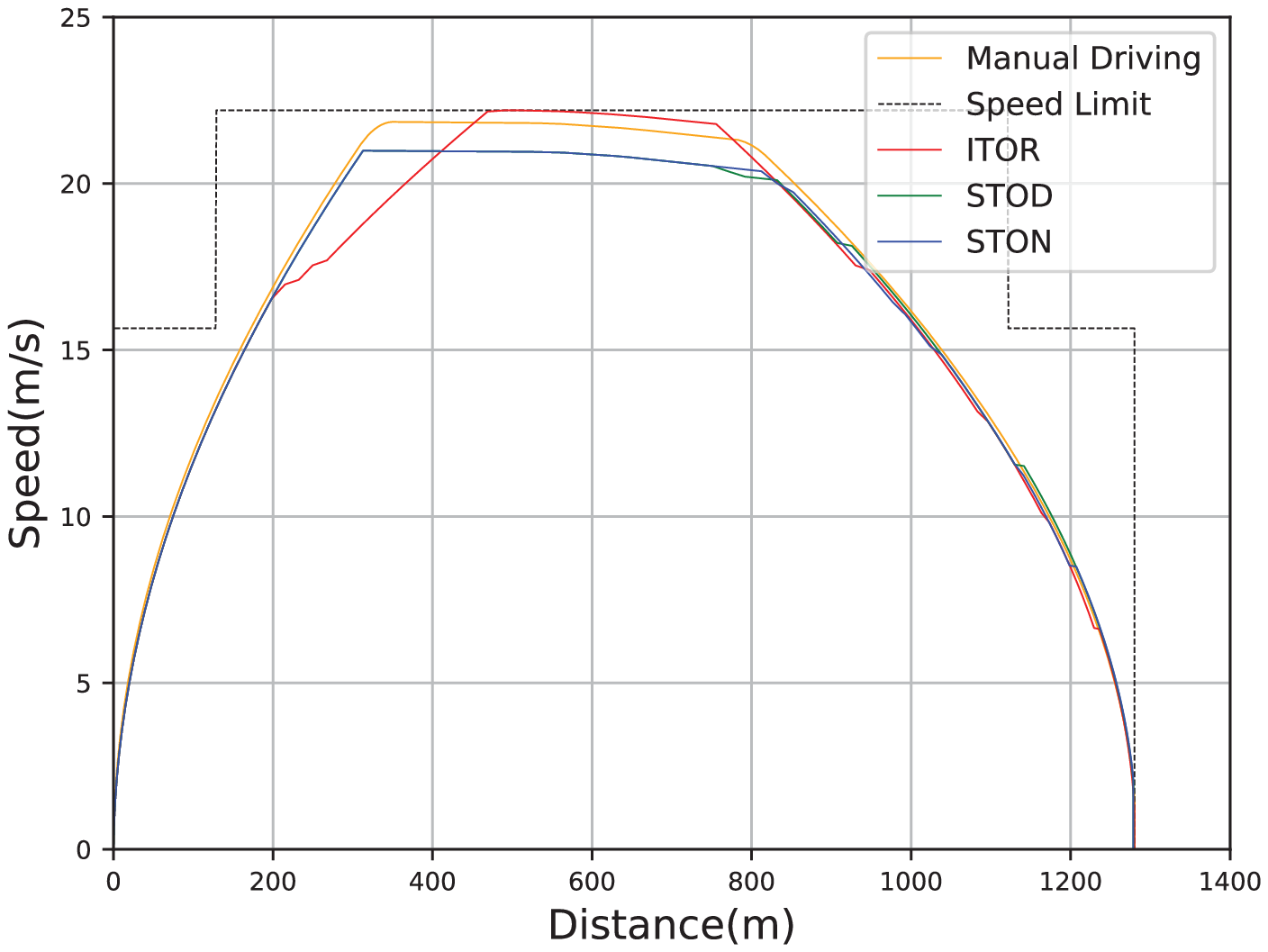}\\
	\end{center}
	\caption{Comparison of speed distance profile with $95s$ trip time.}
	\label{fig: 95}
\end{figure}

\par 
Fig. \ref{fig: 95} presents speed distance profiles for $95s$ trip time of the four models. Similar to the case with $101s$ trip time, the speed-distance profile of ITOR has the highest maximum speed, which indicates that ITOR has the largest energy consumption. The speed profiles of STOD and STON are very similar, as they have the same maximum speed $20.99m/s$. Compared with the speed-distance profile under $101s$ planning trip time, they have a higher maximum speed and short coasting distance which indicates higher energy consumption and worse passenger comfort. 

\par We can learn from the Table. \ref{tab2} that compared with the manual driving, ITOR costs $5.2\%$ of energy more than manual driving; STOD costs $8.7\%$ of energy less than the manual driving and STON costs $8.8\%$ of energy less than the manual driving. In the aspect of the riding comfort for four approaches, manual driving has the largest $I_c$, while ITOR, STOD, and STON have similar $I_c$ which is much smaller than that of the manually driving method. Among them, STON has the best $I_c$ and $I_e$ with the trip time $95s$.

\begin{figure}
	\begin{center}
		\centering
		\scriptsize
		\includegraphics[width=3.6in]{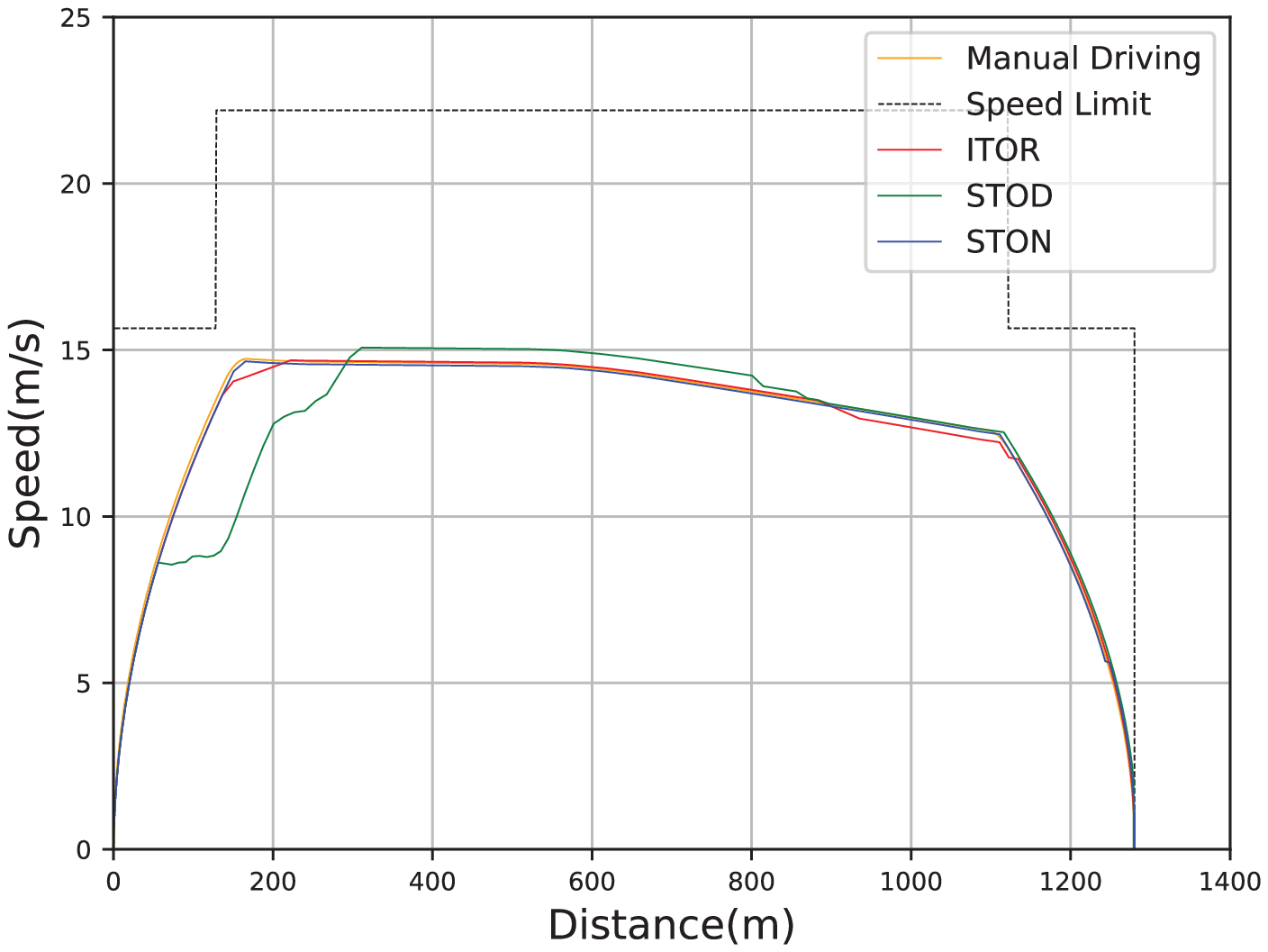}\\
	\end{center}
	\caption{Comparison of speed distance profile with $115s$ trip time.}
	\label{fig: 115}
\end{figure}

\par Fig. \ref{fig: 115} presents the comparison of speed distance profiles with $115s$ trip time. The maximum speed of the manually driving is $14.73m/s$;  the maximum speed of the ITOR is $14.69m/s$; the maximum speed of the STOD is $15.07m/s$ and the maximum speed of the STON is $14.65m/s$. We can learn that the STOD has the highest maximum speed than the other three methods. Compared with their speed distance profiles under $95s$ and $101s$ planning trip time, their maximum speeds are much lower than that of previous cases which denotes that they have lower average speed and lower energy consumption. 

\par We can learn from the Table. \ref{tab2} that in terms of punctuality and safety, all four models satisfy the requirements. As to energy consumption, the STOD has the highest energy consumption. Compared with the manually driving,  ITOR costs $0.5\%$ of energy more than manual driving; STOD costs $6.1\%$ of energy more than the manual driving and STON costs $1.5\%$ of energy less than the manual driving. The comfort evaluation index of STOD is higher than that of the other three models, as its acceleration changes several times during the accelerating phase. This time, the $I_c$ of ITOR is lower than that of STON while the $I_e$ of STON is lower than that of ITOR.

\par Through the analysis above, we can conclude that ITOR, STON, and STOD can produce rational control strategy and satisfy all requirements when the planning trip time is changed, thus the flexibility of ITOR, STOD, and STON is proved.

\subsection{Case 3}
\begin{figure}
	\begin{center}
		\centering
		\scriptsize
		\includegraphics[width=3.6in]{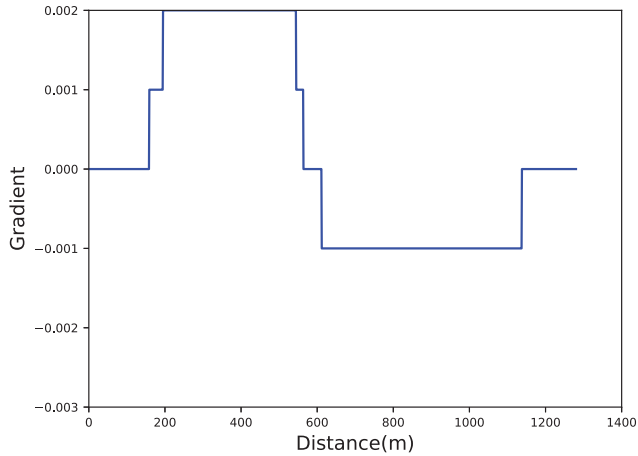}\\
	\end{center}
	\caption{New gradient condition between Rongjing East Street
		station and Wanyuan Street station.}
	\label{fig: NEWGR}
\end{figure}

To test the robustness of STOD and STON, we will change the gradient condition in the same railway section of YLBS. Even though in most cases, the gradient condition is stable, other factors like the wet weather and the railway aging are able to change the resistance condition of the railway. In this case, through changing the gradient condition, we can simulate the situation of the changing resistance, which can be used to test the robustness of STOD and STON. The new gradient condition is shown in Fig. \ref{fig: NEWGR}.

\par
\begin{figure}
	\begin{center}
		\centering
		\scriptsize
		\includegraphics[width=3.6in]{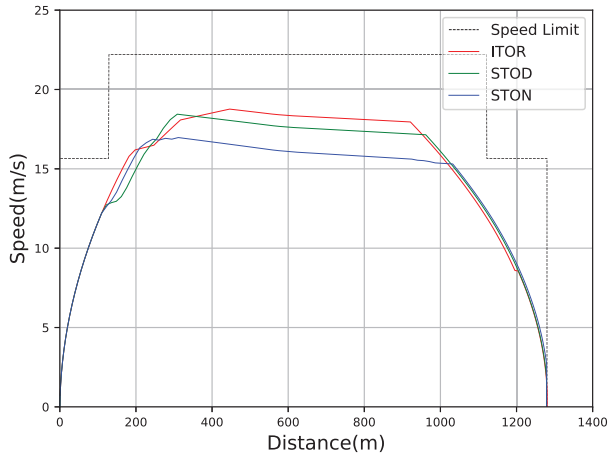}\\
	\end{center}
	\caption{Comparison of speed distance profile with new gradient condition.}
	\label{fig: gr}
\end{figure}

 Fig. \ref{fig: gr} presents the comparison of speed distance profiles with the new gradient condition. The maximum speed of the ITOR is $18.76m/s$; the maximum speed of the STOD is $18.09m/s$ and the maximum speed of the STON is $16.97m/s$. We can learn that ITOR has the highest maximum speed than the other two methods. 

\par We can learn from the Table. \ref{tab2} that all three models satisfy the requirement of punctuality and safety. With the same planning trip time, the energy consumption of ITOR and STOD are a little higher than that in \emph{Case 1}, as new gradient condition increases the resistance of the section where subway accelerates and decreases the resistance of the section where subway decelerates. The speed profile given by the STON has lower energy consumption, while the main reason is that the running time of STON is $103s$ rather than $101s$, which is $2s$ later than expected. However, according to the definition of punctuality which indicates that time errors inferior to $3s$ are allowable, the STON still provides a good result. The comfort evaluation index of ITOR and STOD are similar to that in \emph{Case 1}, whereas the comfort evaluation index of STON is lower than that in \emph{Case 1}, as both the accelerating phrase and the decelerating phase of STON speed distance profile, in this case, are smoother than that in \emph{Case 1}. We can learn from the result listed above, that ITOR, STOD, and STON are capable to provide satisfactory results even when the resistance condition varies, hence the robustness of ITOR, STOD, and STON are verified.

\section{Conclusion}
To build an intelligent train operation model that can deal with the control task for continuous action of the subway, we propose two algorithms, STOD and STON which integrate the expert knowledge of experienced drivers with reinforcement learning methods. Firstly, we collect enough driving data of experienced drivers, from whom we extract expert knowledge and build inference methods. Then we integrate expert knowledge rules and inference methods into reinforcement learning algorithms, 
DDPG and NAF. Finally, three case studies based on YLBS are used to illustrate the effectiveness, the flexibility and the robustness of STOD and STON. The performance of the proposed STOD and STON is compared with the performance of the existing ATO algorithm. The result shows that STOD and STON perform better than manually driving and existing ATO algorithm. Two proposed models own certain flexibility and robustness, which allows them to deal with the variability of subway operation tasks. Moreover, STON generally performs better than STOD in the comparisons of simulation results for three cases listed above.
\par Besides the feature of dealing with control problems for continuous action, STOD and STON also make use of the expert knowledge and inference methods, which largely increases the stability of algorithms. In addition, STOD and STON can meet multiple objectives of train operation and realize model-free train operation control. 
\par However, despite these advantages mentioned above, proposed models are still improvable. For example, the flexibility of STOD and STON is limited. If a great change in the planning trip time is made, STOD and STON cannot generate a desirable control strategy. Moreover, it is hard to apply these models to high-speed train cases with long-distance and complicated speed limits between two successive stations, which decreases the convergence speed of the learning process of algorithms and also increases the instability of models. Our future research will focus on these aspects.

\section{Acknowledgement}
The first author really appreciates the support and the company from his colleagues, Peng Jiang, and Zeyu Zhao, in the Department of Automation of Tsinghua University.

\bibliographystyle{IEEEtran}%
\bibliography{Main}

\begin{thebibliography}{10}
\providecommand{\url}[1]{#1}
\csname url@samestyle\endcsname
\providecommand{\newblock}{\relax}
\providecommand{\bibinfo}[2]{#2}
\providecommand{\BIBentrySTDinterwordspacing}{\spaceskip=0pt\relax}
\providecommand{\BIBentryALTinterwordstretchfactor}{4}
\providecommand{\BIBentryALTinterwordspacing}{\spaceskip=\fontdimen2\font plus
\BIBentryALTinterwordstretchfactor\fontdimen3\font minus
  \fontdimen4\font\relax}
\providecommand{\BIBforeignlanguage}[2]{{%
\expandafter\ifx\csname l@#1\endcsname\relax
\typeout{** WARNING: IEEEtran.bst: No hyphenation pattern has been}%
\typeout{** loaded for the language `#1'. Using the pattern for}%
\typeout{** the default language instead.}%
\else
\language=\csname l@#1\endcsname
\fi
#2}}
\providecommand{\BIBdecl}{\relax}
\BIBdecl

\bibitem{yin2016smart}
J.~Yin, D.~Chen, and Y.~Li, ``Smart train operation algorithms based on expert
  knowledge and ensemble cart for the electric locomotive,''
  \emph{Knowledge-Based Systems}, vol.~92, pp. 78--91, 2016.

\bibitem{corman2015review}
F.~Corman and L.~Meng, ``A review of online dynamic models and algorithms for
  railway traffic management,'' \emph{IEEE Transactions on Intelligent
  Transportation Systems}, vol.~16, no.~3, pp. 1274--1284, 2015.

\bibitem{albrecht2013energy}
A.~R. Albrecht, P.~G. Howlett, P.~J. Pudney, and X.~Vu, ``Energy-efficient
  train control: from local convexity to global optimization and uniqueness,''
  \emph{Automatica}, vol.~49, no.~10, pp. 3072--3078, 2013.

\bibitem{wang2015efficient}
Y.~Wang, B.~Ning, T.~Tang, T.~J. Van Den~Boom, and B.~De~Schutter, ``Efficient
  real-time train scheduling for urban rail transit systems using iterative
  convex programming,'' \emph{IEEE Transactions on Intelligent Transportation
  Systems}, vol.~16, no.~6, pp. 3337--3352, 2015.

\bibitem{yang2014two}
X.~Yang, B.~Ning, X.~Li, and T.~Tang, ``A two-objective timetable optimization
  model in subway systems,'' \emph{IEEE Transactions on Intelligent
  Transportation Systems}, vol.~15, no.~5, pp. 1913--1921, 2014.

\bibitem{shangguan2015multiobjective}
W.~ShangGuan, X.-H. Yan, B.-G. Cai, and J.~Wang, ``Multiobjective optimization
  for train speed trajectory in ctcs high-speed railway with hybrid
  evolutionary algorithm,'' \emph{IEEE Transactions on Intelligent
  Transportation Systems}, vol.~16, no.~4, pp. 2215--2225, 2015.

\bibitem{khmelnitsky2000optimal}
E.~Khmelnitsky, ``On an optimal control problem of train operation,''
  \emph{IEEE Transactions on Automatic Control}, vol.~45, no.~7, pp.
  1257--1266, 2000.

\bibitem{accikbacs2008coasting}
S.~A{\c{c}}{\i}kba{\c{s}} and M.~S{\"o}ylemez, ``Coasting point optimisation
  for mass rail transit lines using artificial neural networks and genetic
  algorithms,'' \emph{IET Electric Power Applications}, vol.~2, no.~3, pp.
  172--182, 2008.

\bibitem{yang2012optimizing}
L.~Yang, K.~Li, Z.~Gao, and X.~Li, ``Optimizing trains movement on a railway
  network,'' \emph{Omega}, vol.~40, no.~5, pp. 619--633, 2012.

\bibitem{yin2014intelligent}
J.~Yin, D.~Chen, and L.~Li, ``Intelligent train operation algorithms for subway
  by expert system and reinforcement learning,'' \emph{IEEE Transactions on
  Intelligent Transportation Systems}, vol.~15, no.~6, pp. 2561--2571, 2014.

\bibitem{liu2013high}
W.~Liu, J.~Han, and X.~Lu, ``A high speed railway control system based on the
  fuzzy control method,'' \emph{Expert Systems with Applications}, vol.~40,
  no.~15, pp. 6115--6124, 2013.

\bibitem{wu2015trajectory}
T.-S. Wu, M.~Karkoub, C.-C. Weng, and W.-S. Yu, ``Trajectory tracking for
  uncertainty time delayed-state self-balancing train vehicles using
  observer-based adaptive fuzzy control,'' \emph{Information Sciences}, vol.
  324, pp. 1--22, 2015.

\bibitem{gu2014energy}
Q.~Gu, T.~Tang, F.~Cao, and Y.-d. Song, ``Energy-efficient train operation in
  urban rail transit using real-time traffic information,'' \emph{IEEE
  Transactions on Intelligent Transportation Systems}, vol.~15, no.~3, pp.
  1216--1233, 2014.

\bibitem{li2014robust}
S.~Li, L.~Yang, K.~Li, and Z.~Gao, ``Robust sampled-data cruise control
  scheduling of high speed train,'' \emph{Transportation Research Part C:
  Emerging Technologies}, vol.~46, pp. 274--283, 2014.

\bibitem{liu2003energy}
R.~R. Liu and I.~M. Golovitcher, ``Energy-efficient operation of rail
  vehicles,'' \emph{Transportation Research Part A: Policy and Practice},
  vol.~37, no.~10, pp. 917--932, 2003.

\bibitem{rochard2000review}
B.~P. Rochard and F.~Schmid, ``A review of methods to measure and calculate
  train resistances,'' \emph{Proceedings of the Institution of Mechanical
  Engineers, Part F: Journal of Rail and Rapid Transit}, vol. 214, no.~4, pp.
  185--199, 2000.

\bibitem{wang2013optimal}
Y.~Wang, B.~De~Schutter, T.~J. van~den Boom, and B.~Ning, ``Optimal trajectory
  planning for trains--a pseudospectral method and a mixed integer linear
  programming approach,'' \emph{Transportation Research Part C: Emerging
  Technologies}, vol.~29, pp. 97--114, 2013.

\bibitem{song2011computationally}
Q.~Song, Y.-d. Song, T.~Tang, and B.~Ning, ``Computationally inexpensive
  tracking control of high-speed trains with traction/braking saturation,''
  \emph{IEEE Transactions on Intelligent Transportation Systems}, vol.~12,
  no.~4, pp. 1116--1125, 2011.

\bibitem{chen2013online}
D.~Chen, R.~Chen, Y.~Li, and T.~Tang, ``Online learning algorithms for train
  automatic stop control using precise location data of balises,'' \emph{IEEE
  Transactions on Intelligent Transportation Systems}, vol.~14, no.~3, pp.
  1526--1535, 2013.

\bibitem{zheng2009modeling}
W.~ZHENG and H.~Xu, ``Modeling and safety analysis of maglev train over-speed
  protection based on stochastic petri nets,'' \emph{Journal of the China
  Railway Society}, vol.~31, no.~4, pp. 59--64, 2009.

\bibitem{olsson2004influencing}
N.~O. Olsson and H.~Haugland, ``Influencing factors on train
  punctuality—results from some norwegian studies,'' \emph{Transport Policy},
  vol.~11, no.~4, pp. 387--397, 2004.

\bibitem{miyatake2010optimization}
M.~Miyatake and H.~Ko, ``Optimization of train speed profile for minimum energy
  consumption,'' \emph{IEEJ Transactions on Electrical and Electronic
  Engineering}, vol.~5, no.~3, pp. 263--269, 2010.

\bibitem{karakasis2005factorial}
K.~Karakasis, D.~Skarlatos, and T.~Zakinthinos, ``A factorial analysis for the
  determination of an optimal train speed with a desired ride comfort,''
  \emph{Applied Acoustics}, vol.~66, no.~10, pp. 1121--1134, 2005.

\bibitem{cheng2017intelligent}
R.~Cheng, D.~Chen, B.~Cheng, and S.~Zheng, ``Intelligent driving methods based
  on expert knowledge and online optimization for high-speed trains,''
  \emph{Expert Systems with Applications}, vol.~87, pp. 228--239, 2017.

\bibitem{powell2015passenger}
J.~Powell and R.~Palac{\'\i}n, ``Passenger stability within moving railway
  vehicles: limits on maximum longitudinal acceleration,'' \emph{Urban Rail
  Transit}, vol.~1, no.~2, pp. 95--103, 2015.

\bibitem{hoberock1977survey}
L.~L. Hoberock, ``A survey of longitudinal acceleration comfort studies in
  ground transportation vehicles,'' \emph{Journal of Dynamic Systems,
  Measurement, and Control}, vol.~99, no.~2, pp. 76--84, 1977.

\bibitem{ruelens2015learning}
F.~Ruelens, S.~Iacovella, B.~J. Claessens, and R.~Belmans, ``Learning agent for
  a heat-pump thermostat with a set-back strategy using model-free
  reinforcement learning,'' \emph{Energies}, vol.~8, no.~8, pp. 8300--8318,
  2015.

\bibitem{murrell1997survey}
S.~Murrell and R.~T. Plant, ``A survey of tools for the validation and
  verification of knowledge-based systems: 1985--19951,'' \emph{Decision
  Support Systems}, vol.~21, no.~4, pp. 307--323, 1997.

\bibitem{zhao2017integrated}
N.~Zhao, C.~Roberts, S.~Hillmansen, Z.~Tian, P.~Weston, and L.~Chen, ``An
  integrated metro operation optimization to minimize energy consumption,''
  \emph{Transportation Research Part C: Emerging Technologies}, vol.~75, pp.
  168--182, 2017.

\bibitem{sutton1998introduction}
R.~S. Sutton and A.~G. Barto, \emph{Introduction to reinforcement
  learning}.\hskip 1em plus 0.5em minus 0.4em\relax MIT press Cambridge, 1998,
  vol. 135.

\bibitem{kaelbling1996reinforcement}
L.~P. Kaelbling, M.~L. Littman, and A.~W. Moore, ``Reinforcement learning: A
  survey,'' \emph{Journal of Artificial Rntelligence Research}, vol.~4, pp.
  237--285, 1996.

\bibitem{abbeel2007application}
P.~Abbeel, A.~Coates, M.~Quigley, and A.~Y. Ng, ``An application of
  reinforcement learning to aerobatic helicopter flight,'' in \emph{Advances in
  Neural Information Processing Systems}, 2007, pp. 1--8.

\bibitem{duan2007application}
Y.~Duan, Q.~Liu, and X.~Xu, ``Application of reinforcement learning in robot
  soccer,'' \emph{Engineering Applications of Artificial Intelligence},
  vol.~20, no.~7, pp. 936--950, 2007.

\bibitem{ernst2004power}
D.~Ernst, M.~Glavic, and L.~Wehenkel, ``Power systems stability control:
  reinforcement learning framework,'' \emph{IEEE Transactions on Power
  Systems}, vol.~19, no.~1, pp. 427--435, 2004.

\bibitem{lillicrap2015continuous}
T.~P. Lillicrap, J.~J. Hunt, A.~Pritzel, N.~Heess, T.~Erez, Y.~Tassa,
  D.~Silver, and D.~Wierstra, ``Continuous control with deep reinforcement
  learning,'' \emph{arXiv preprint arXiv:1509.02971}, 2015.

\bibitem{mnih2013playing}
V.~Mnih, K.~Kavukcuoglu, D.~Silver, A.~Graves, I.~Antonoglou, D.~Wierstra, and
  M.~Riedmiller, ``Playing atari with deep reinforcement learning,''
  \emph{arXiv preprint arXiv:1312.5602}, 2013.

\bibitem{mnih2015human}
V.~Mnih, K.~Kavukcuoglu, D.~Silver, A.~A. Rusu, J.~Veness, M.~G. Bellemare,
  A.~Graves, M.~Riedmiller, A.~K. Fidjeland, G.~Ostrovski \emph{et~al.},
  ``Human-level control through deep reinforcement learning,'' \emph{Nature},
  vol. 518, no. 7540, p. 529, 2015.

\bibitem{gu2016continuous}
S.~Gu, T.~Lillicrap, I.~Sutskever, and S.~Levine, ``Continuous deep q-learning
  with model-based acceleration,'' in \emph{International Conference on Machine
  Learning}, 2016, pp. 2829--2838.

\end{thebibliography}

\begin{IEEEbiography}[{\includegraphics[width=1in,height=1.25in,clip,keepaspectratio]{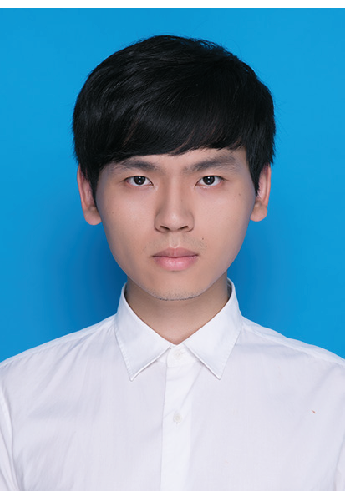}}]{Kaichen Zhou}
	received the Master degree in Machine Learning in the Department of Computing at Imperial College London. He is currently pursuing the P.h.D degree in Computer Science in the Department of Computer Science at University of Oxford.
	\par His research interests include deep learning and reinforcement learning. 
\end{IEEEbiography}
\begin{IEEEbiography}[{\includegraphics[width=1in,height=1.25in,clip,keepaspectratio]{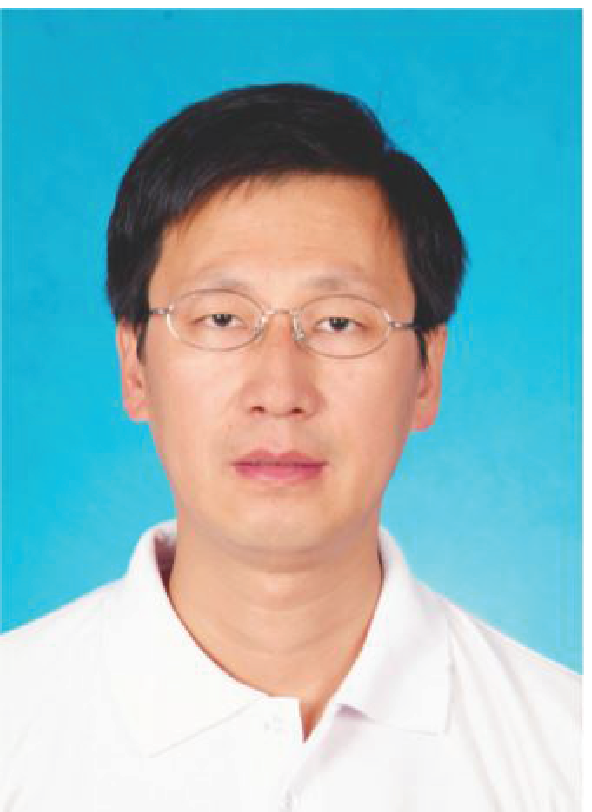}}]{Shiji Song}
	received the Ph.D. degree in Mathematics from the Department of Mathematics, Harbin Institute of Technology, Harbin, China, in 1996. 
	\par He is currently a Professor with the Department of Automation, Tsinghua University, Beijing, China. His current research interests include system modeling, control and optimization, computational intelligence, and pattern recognition.
\end{IEEEbiography}
\begin{IEEEbiography}[{\includegraphics[width=1in,height=1.25in,clip,keepaspectratio]{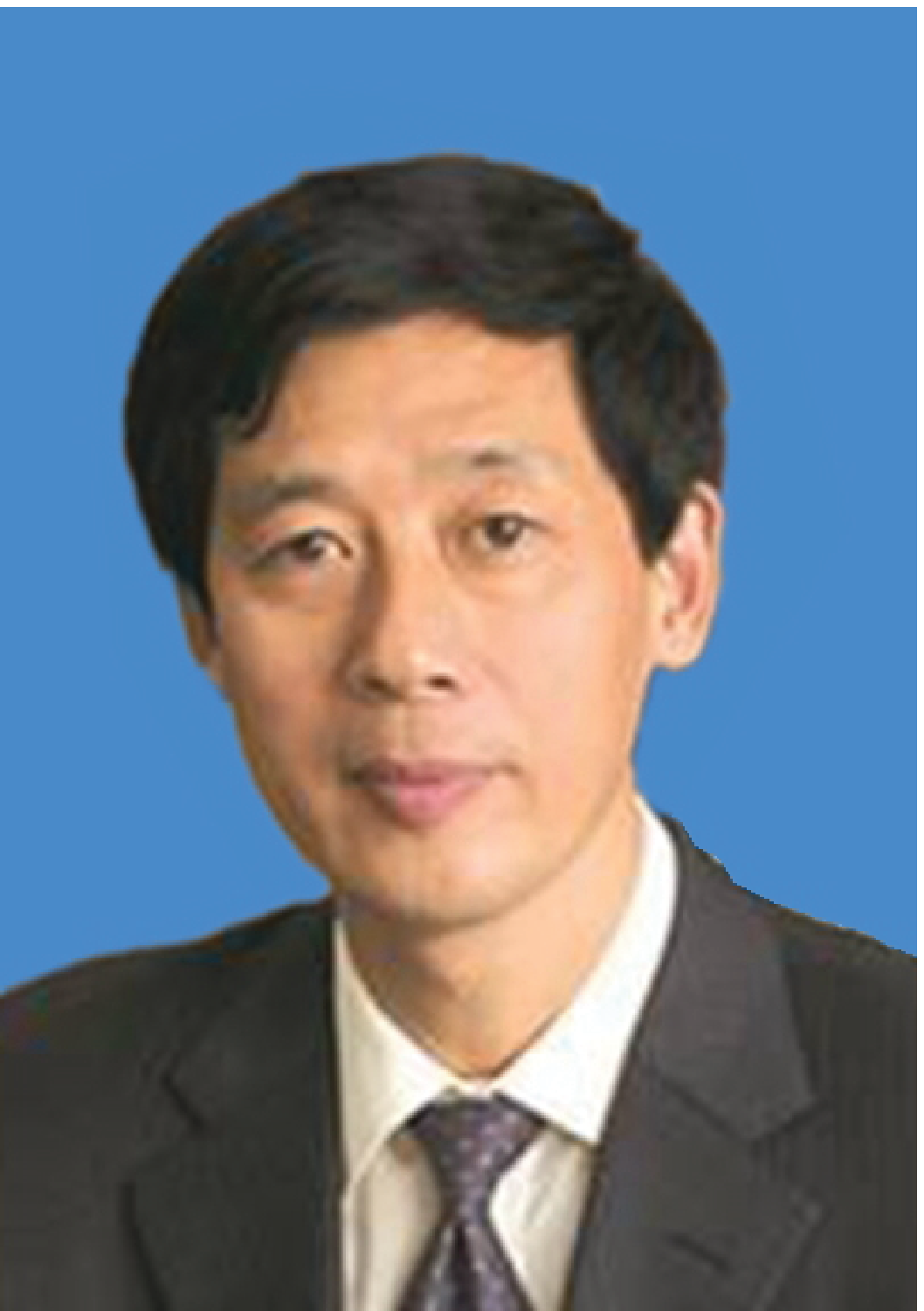}}]{Anke Xue}
	received the Ph.D. degree in Automation from the College of Control Science and Engineering, Zhejiang University, Hangzhou, China, in 1997.
	\par He is currently a Professor and the President of Hangzhou Dianzi University, Hangzhou. His research interests include robust control theory and applications.
\end{IEEEbiography}
\begin{IEEEbiography}[{\includegraphics[width=1in,height=1.25in,clip,keepaspectratio]{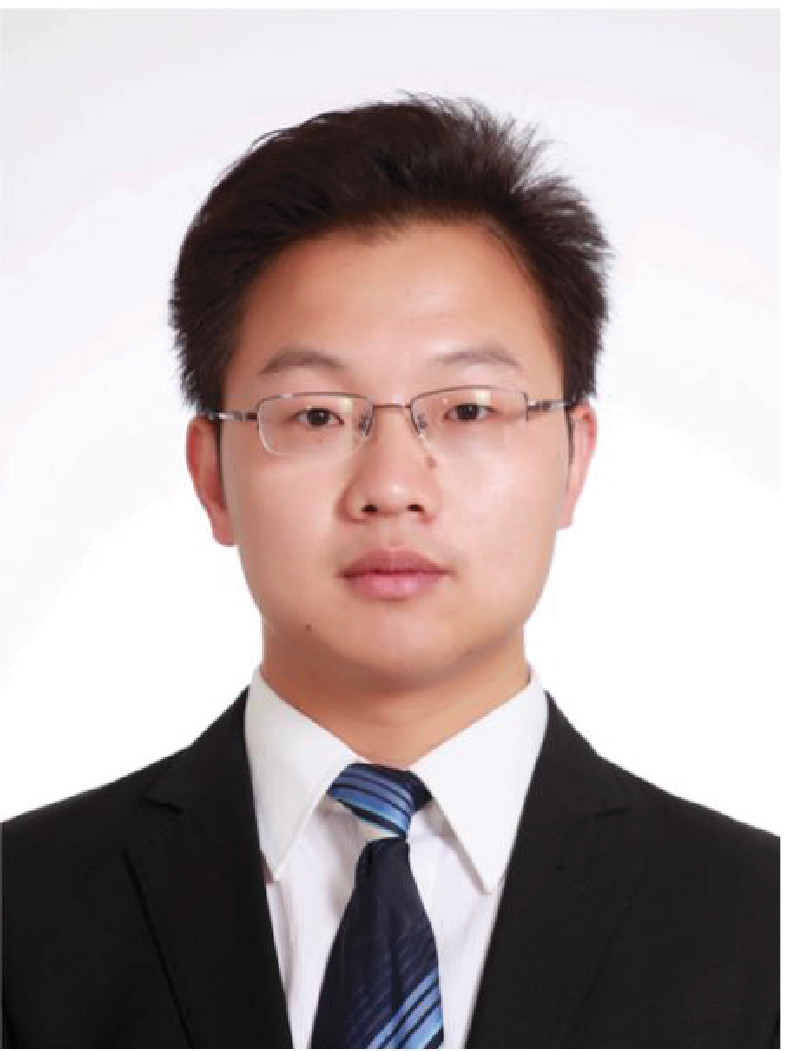}}]{Keyou You}
	received the B.S. degree in Statistical Science from Sun Yat-sen University, Guangzhou, China, in 2007 and the Ph.D. degree in Electrical and Electronic Engineering from Nanyang Technological University (NTU), Singapore, in 2012.
	He is now an Associate Professor in the Department of Automation, Tsinghua University, Beijing, China. He held visiting positions at Politecnicodi Torino, Hong Kong University of Science and Technology, University of Melbourne and etc. \par His current research interests include networked control systems, parallel networked algorithms, and their applications. 
\end{IEEEbiography}
\begin{IEEEbiography}[{\includegraphics[width=1in,height=1.25in,clip,keepaspectratio]{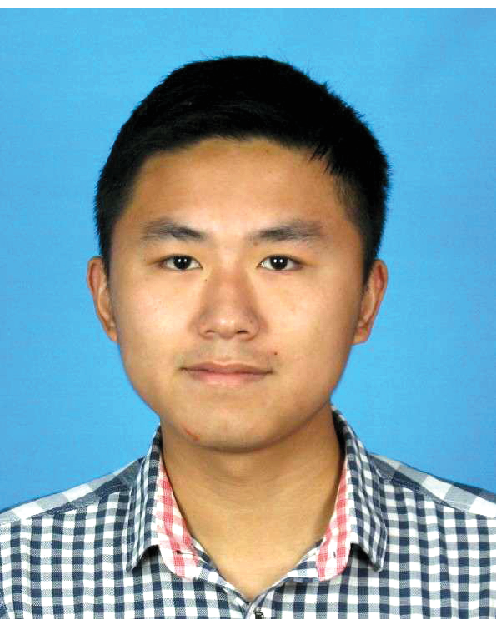}}]{Hu Wu}
	Hui Wu received the B.S. degree in Automation from the Department of Automation, Tsinghua University, Beijing, China, in 2014, where he is currently pursuing the Ph.D. degree in control science and engineering with the Department of Automation, Institute of System Integration in Tsinghua University.
	\par His current research interests include reinforcement learning and robot control, especially in continuous control for underwater vehicles.
\end{IEEEbiography}

\end{document}